\documentclass[11pt,preprint]{aastex}

\begin{document}

\title{Solar Eruption and Local Magnetic Parameters}
\author{Jeongwoo Lee\altaffilmark{1,2}, Chang Liu\altaffilmark{3}, Ju Jing\altaffilmark{3}, Jongchul Chae\altaffilmark{1}}
\affil{1. Department of Physics and Astronomy, Seoul National University, Seoul 08826, Korea}
\affil{2. Institute for Space-Earth Environmental Research, Nagoya University, Aichi 464-8601, Japan}
\affil{3. Space Weather Research Laboratory, New Jersey Institute of Technology, Newark, NJ 07102, U.S.A}

\begin{abstract}
It is now a common practice to use local magnetic parameters such as magnetic decay index for explaining solar eruptions from active regions, but there can be an alternative view that the global properties of the source region should be counted as a more important factor. We discuss this issue based on {\it Solar Dynamics Observatory} (SDO) observations of  the three successive eruptions within 1.5 hours from the NOAA active region 11444 and the magnetic parameters calculated using the nonlinear force-free field (NLFFF) model. Two violent eruptions occurred in the regions with relatively high magnetic twist number (0.5--1.5) and high decay index (0.9--1.1) at the nominal height of the filament (12$''$) and otherwise a mild eruption occurred, which supports the local parameter paradigm. Our main point is that the time sequence of the eruptions did not go with these parameters. It is argued that an additional factor, in the form of stabilizing force, should operate to determine the onset of the first eruption and temporal behaviors of subsequent eruptions.
As supporting evidence, we report that the heating and fast plasma flow continuing for a timescale of an hour was the direct cause for the first eruption, and that the unidirectional propagation of the disturbance determined the timing of subsequent eruptions. Both of these factors are associated with the overall magnetic structure rather than local magnetic properties of the active region.

\end{abstract}


\section{INTRODUCTION}

Solar eruptions play a crucial role in space environment, and it is desirable to understand how they are triggered (e.g., Forbes 2000). Although the triggering mechanisms for solar eruptions is still debated, many theoretical models present the local magnetic parameters as the criterion for the specific type of magnetic instability.
For kink instability, we require that the flux tube should be twisted by more than 1.75 turns (Berger \& Prior 2006, T\"or\"ok et al. 2004). The torus instability, now regarded as the main driver of eruption (D\'emoulin \& Aulanier 2010) requires that the overlying magnetic field should decrease with height faster than a critical value. Commonly used value for the critical decay index for the torus instability is 1.5 based on the analytical study by Bateman (1978) and Kliem \& T\"or\"ok (2006). However, an extensive MHD simulation study found  the critical value lying in the range of 1.3--1.5 (Zuccarello, 2015, 2016). Another analytical study found a critical range between 1.1 and 1.3, and it depends on the magnetic field and electric current configuration (Demoulin \& Aulanier 2010).  Many studies have been performed to evaluate whether these local parameters are viable in explaining solar eruptions (Sun et al. 2015, Kusano et al. 2012, 2004, Guo et al. 2010, Fan 2005, Lippiello et al. 2008, Chen \& Shibata 2000).

There can, however, be an alternative view that the local parameter may not explain all aspects of the onset of solar eruptions. Myers et al. (2015) found, based on a laboratory experiment, an additional eruption criterion is needed because in the presence of strong guide magnetic field a new tension force arises to prevent the flux rope from kinking. This guide magnetic field may work as another local parameter, but can be a global property depending on its extent.
Hudson (2011) pointed out the importance of the global quantities in solar flares. For the global properties the energetics of the flare radiation, coronal mass ejections (CMEs), and waves were discussed.  Ultimately we should be able to relate the observable global quantities to the fundamental quantities such as thresholds in magnetic free energy and magnetic helicity.

In this Letter we challenge the prevailing idea of using those local parameters for explaining solar eruptions, focusing on two ideal MHD instabilities, torus and kink instabilities, for which we can evaluate the onset criteria based on observable quantities together with the coronal field extrapolation techniques. This precludes the tearing mode instability   (Furth et al., 1963), which is local in nature, but involves microscopic processes hardly accessible from observations.
It is generally difficult to associate an eruption event to a global property when a single eruption occurs in a local region known to be already unstable. We thus study a series of eruptions from an AR so that we can check the relative timing, location, and strength of each eruption against the local parameters. Under this strategy we do not directly address specific global parameters, but may advocate (dismiss) the  local-parameter paradigm based on success (failure) of the test.
We describe the data in \S 2, and the eruptions in \S 3. We then investigate the local magnetic parameters in \S4, and conclude in \S5.

\section{Data}

The target selected for this study is the NOAA AR 11444, which produced three eruptions on March 27 in 2012. According to a preflare vector magnetogram of this AR from the {\it Helioseismic and Magnetic Imager} (HMI; Schou et al. 2012) on board {\it Solar Dynamics Observatory} (SDO; Pesnell et al. 2012) it has the positive polarity magnetic fields in the center and the negative polarity fields around so that the polarity inversion line (PIL) takes a quasi-circular form. The eruption characteristics associated with a quasi-circular PIL structure has been presented in Lee et al. (2016). In this Letter we concern ourselves with the successive eruptions occurred along the PIL, which provides an excellent opportunity to check  the temporal and spatial properties of each eruption against  the physical parameters around the AR.
We use the EUV images obtained from the {\it Atmospheric Imaging Assembly} (AIA; Lemen et al. 2012) on board the SDO, in particular, the EUV images obtained at the wavelength of 171 \AA\ and 304 \AA\ which correspond to the corona and chromosphere of the Sun, respectively. These images are obtained with a cadence of 12 sec and sampling of 0.6$''$. We have obtained those data at Level-1.0, and used aia$\_$prep.pro routine available in SSW packages to co-align the images from all of the AIA channels to a specified pointing, re-scaled the images to a common plate scale and de-rotate the images to the reference time set to be 02:00 UT, also performing the dark current subtraction and flat-fielding. We finally normalized the exposure time to 1-sec for all images.

\begin{figure}[tbh]
\includegraphics[scale=1]{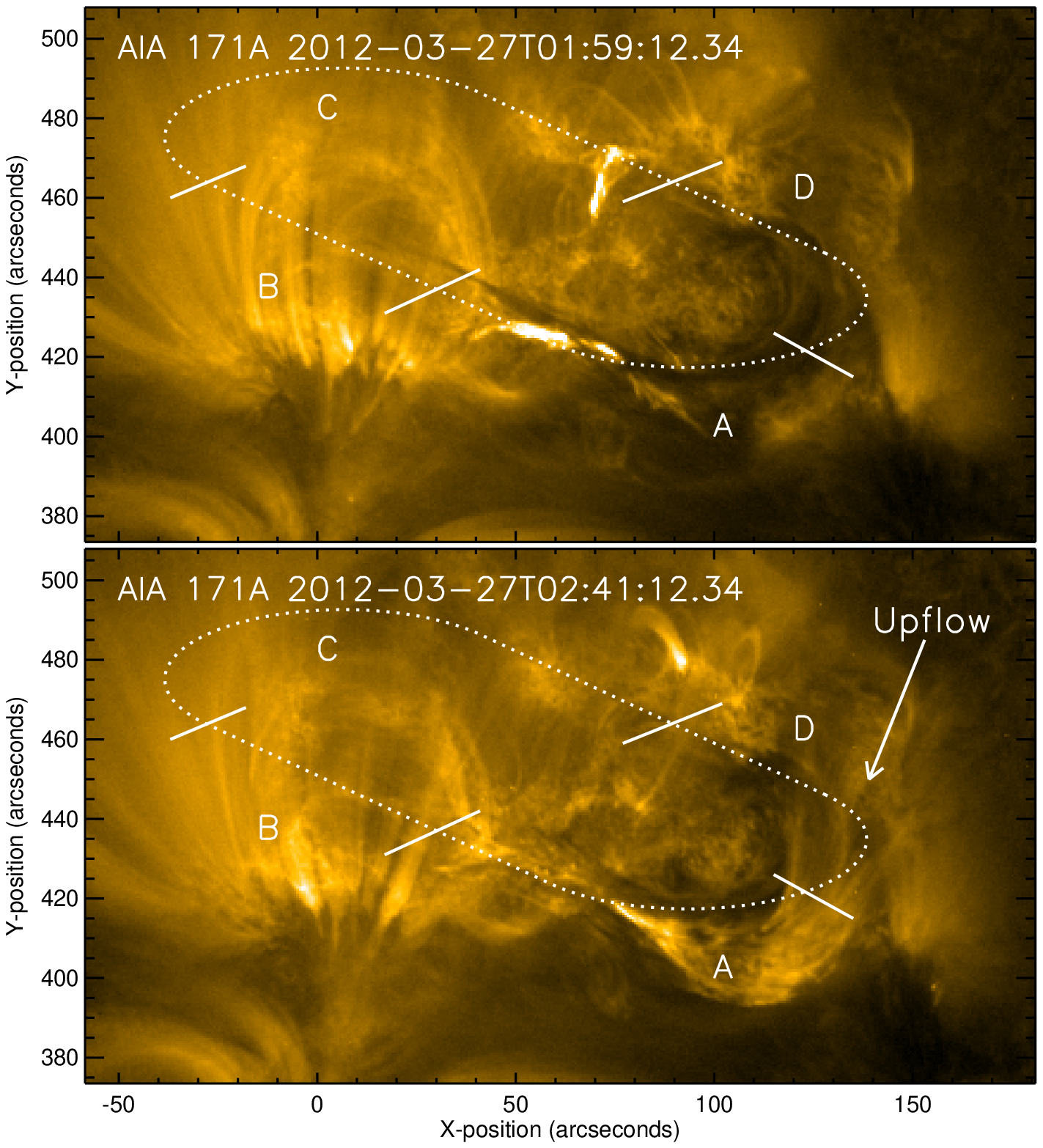}
\caption{The pre-eruption activities in NOAA 11444 on 2012 March 27.  The top panel shows two localized brightenings suggestive of tether-cutting like events. A to D refer to the parts of the filament in which individual eruptions took place. The bottom panel shows fast plasma motions along the filament and upflow along the open field lines, by which the filament became lighter. The dotted oval roughly outlines the filament location. (Movies of these activities observed at 171 \AA\ and 304 \AA\ are available online.)}
\end{figure}

\section{The Eruptions}

Figure 1 highlights two kinds of the pre-eruption activities that occurred in NOAA AR 11444: local brightenings at two places (top panel) and the upward flow along the open field lines (bottom panel). They respectively occurred about 50 min and 10 min before the first eruption ($\sim$02:50 UT) and contributed to the filament activation.
The dotted oval is a rough representation of the filament location.
The main characteristic of this event is the presence of the long filament encircling the AR before the eruptions. It may not be of one single but a few separate filaments.
In view of the eruption activities, it seems appropriate to divide the quasi-circular PIL into four sections denoted as A to D. The filament activation was mostly active in section A. Many sporadic brightenings such as shown in the top panel may represent tether-cutting like reconnection, but none of them immediately led to the lift of the filament. They may have contributed to the heating so that local pressure is enhanced to push materials out of the loop to result in the fast flows observed to run along the filament. Typical speed was 80--100 km s$^{-1}$ as they run a distance of $\sim$40 Mm in  7--8 min.
The first eruption broke out in A, and the filament, F1, is observed to rise nearly uniformly over A (see the movies). {\bf Before the eruption, the significant upward flows and the reduction in the amount of the dark filament material are observed. This implies that the disappearance of the dark filament material was not only due to heating, but the motion of plasma out of the filament leading to mass unloading.}

\begin{figure}[tbh] 
\includegraphics[scale=0.88]{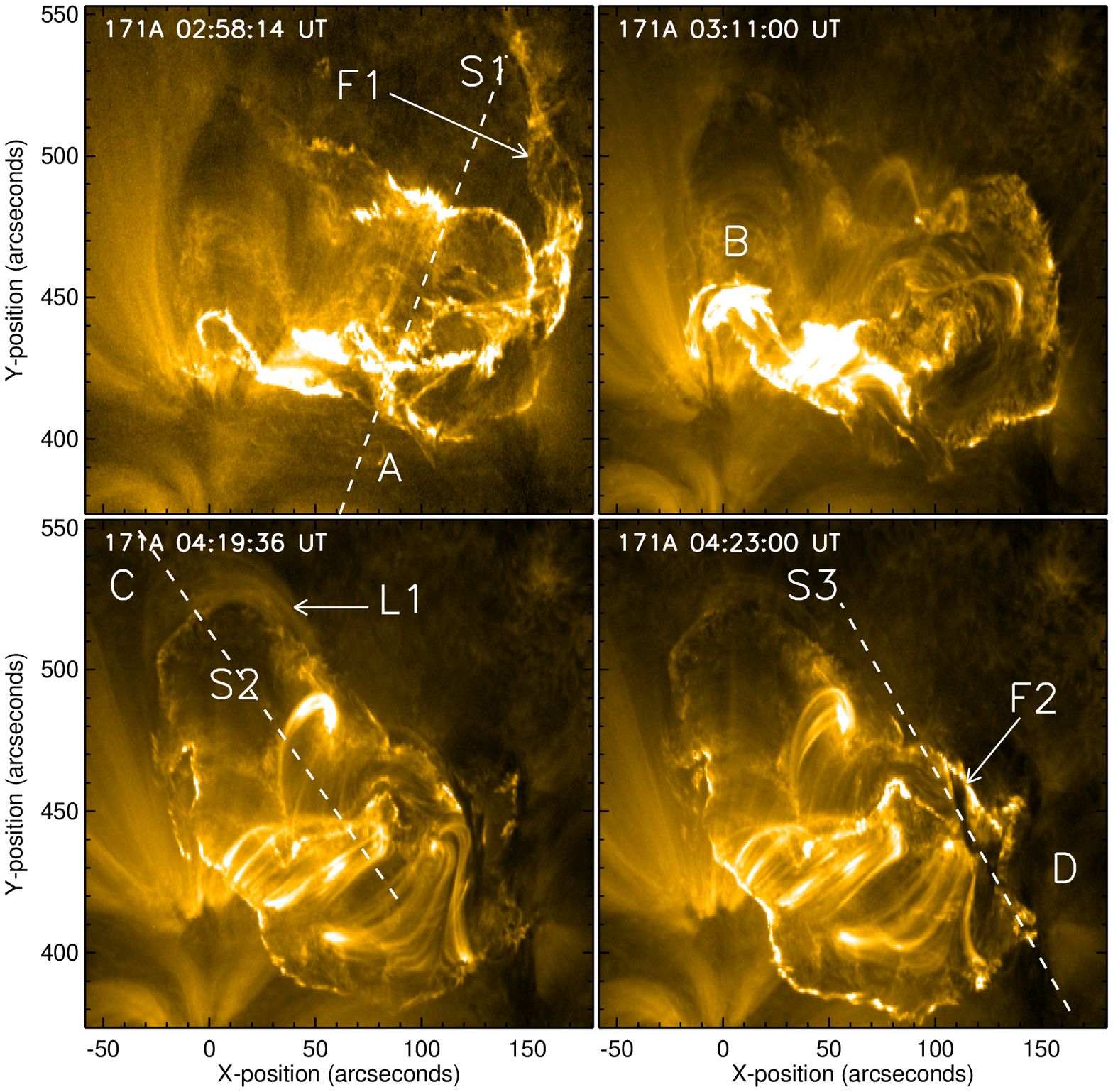}
\caption{The three eruptions seen at AIA 171 \AA\. The first eruption is due to the rise of the filament F1 in region A. The accompanying flare emission is denoted as B. The second eruption took place in the form of an expanding loop, L1, in the eastern region, C. The third eruption occurred in the north-western part, D, due to F2. The dashed lines denote the slits used to derive the time-distance map displayed in Figure 3. }
\end{figure}

Figure 2 shows the three successive eruptions: the rising filament F1 from A during the first eruption, flare emission from the magnetic arcade in B, an expanding magnetic loop, L1, in C, and the third erupting filament, F2, in D. Dashed lines are slits used for measuring the eruption speed.
The first eruption started with the rise of the filament, F1, leaving a two ribbon flare underneath. While the western leg of F1 is well visible even after the eruption (Figure 2$a$), the eastern leg is unclear. Probably,  the post-flare arcade progressively expanding eastward manifests the sequential rise of the eastern leg of F1.

The second eruption occurs in the form of an expanding loop, L1, in C, when the magnetic arcade continues to grow to touch upon the eastern tip of the AR. This eruption is mild. While the loop expansion in C was still in progress, the filament F2 erupted in D. The location of F2 is only slightly displaced from that of F1, and its evolving pattern is very similar to that of F1, in the sense that part of the filament is lifted up to leave flare ribbons underneath. This northern ribbons due to eruption of F2 together with the southern ribbons due to eruption of F1 make the ribbons look quasi-circular. While the initial eruption in A may have been caused by {\bf heating and mass unloading}, and the rest eruptions in C and D occurred sympathetically to each other, and proceeded clockwise along the PIL.

Figure 3 shows the time-distance ($t$--$d$) map constructed from the slits denoted in Figure 2. It is a composite map in which until 03:36 UT we use the slit, S1, on 304 \AA\ images to detect the motion of F1, and S3 after 04:18 to detect that of F2. The middle section was constructed using S2 on 171 \AA\ images, because the loop, L1, is clearly visible only at the coronal temperature. The dashed lines are the linear fits to the eruptive features to measure its speed. We found the eruption speed, $v_1= 360$ km s$^{-1}$ at S1 and $v_2=22$  km s$^{-1}$ at S2. Through S3, two diverging motions, $v_3=270$  km s$^{-1}$ and $v_3'=-$89 km s$^{-1}$, are detected, because the dark material on the rising filament apparently fall apart from each other.
A rough estimate of the viewing angles from the line of sight are 40\degr, 70\degr, and 10\degr, respectively so that these apparent velocities may turn out to be 1100  km s$^{-1}$ for F1,  29 km s$^{-1}$ for L1, 1600 ($-$90)  km s$^{-1}$ for F2, respectively.  The highest eruption speed was therefore produced from D, and the lowest speed was from C.

\begin{figure}[tbh] 
\includegraphics[scale=0.9]{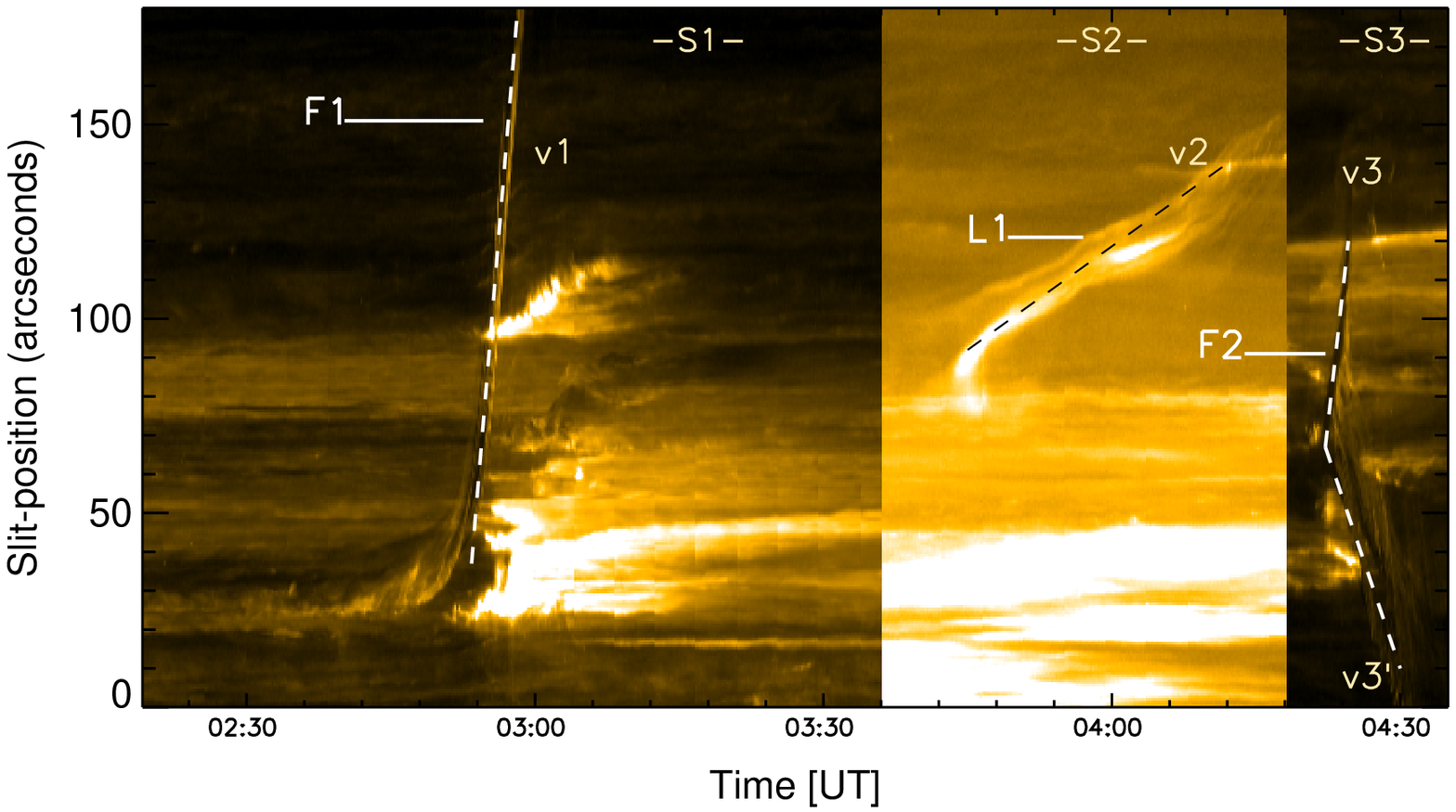}
\caption{A composite time-distance map for the three eruptions. The first section running from 02:19 UT to 03:36 UT is constructed using slit, S1 and the third section starting from 04:18 UT, using S3 on 304 \AA\ images. The middle section is constructed using S2 on 171 \AA\ images. In all cases, the distance is measured northward starting from the southern end of each slit.  Dashed lines denoted by $v_1 ... v_3'$ are the linear fits to the features of the erupting filament for deriving their speeds. }
\end{figure}

\section{Magnetic Parameters}

We use the NLFFF extrapolation code developed by Wiegelmann (2004) to reconstruct the coronal magnetic field from the HMI vector magnetogram at 01:47:17 UT as the boundary condition.
{\bf The force-freeness of the model magnetic field is found satisfactory as the metric $\langle$CW$\sin \theta\rangle$ comes out to be as low as 0.12 (cf., DeRosa et al. 2009).}
The magnetic twist number is calculated as $T=\int (J_{\|} /4\pi B)~dl$ where $B$ is the magnetic field strength,  $J_{\|} $  is the current density running along the field line, and $dl$ is its line element.
The decay index is defined as $n = - d \ln B/d \ln z $, where  $z$ is the vertical height, and $B$ in this case is the potential magnetic field (Fan 2010).
{\bf We estimated the height of the filament as $12''\pm 4''$ based on the length of the filament legs on the AIA 171\AA\ image (Figure 1).}
Figure 4 shows the horizontal distribution of the two quantities at the height of $12''$ (upper panel) and the field lines passing though the PIL at this height (bottom). In the upper panel, the background image represents the distribution of $\alpha$ using the color table of blue ($\alpha >0$), red ($\alpha<0$) and gray ($\alpha=0$).  The thick yellow contour represents the PIL at the height, and the black contours are the decay index in two levels of $n=1.1$ and 1.5. The legend inside the figure shows the scale of $\alpha$ in units of Mm$^{-1}$.

\subsection{Kink Instability}
As shown in the upper panel of Figure 4, the coronal PIL takes a roughly round and closed form, along which the locally enhanced $\alpha$ is found. This shape is associated with the dome-shaped separatrix for this type of AR (Lee et al. 2016). The joint of two different signs of $\alpha$ between A and D may be a notable feature in the context of  the flare model of magnetic helicity annihilation (Romano \& Zuccarello 2011, Kusano et al. 2004).  However, no flare occurred at this location (cf. Shibata \& Magara 2011), and we merely interpret it as indicating F1 being magnetically disconnected from F2.

How much the enhanced $\alpha$ is important to the twist number can be checked from the lower panel of Figure 4 where the field lines are plotted in the interval of 2 arc sec along the PIL at $z=12''$.  Those field lines with relatively high twist number, $T>0.5$, are colored white for distinction.  The open field lines are excluded when counting the twist number. Although magnetic shear and magnetic twist are two different concepts, higher $\alpha$ is often found for the strongly sheared field lines. It is that the more sheared field lines are typically longer than the less-sheared field lines. As a result, the enhanced $\alpha$ is associated with higher $T$. This is why A and D are populated by the highly twisted field lines, whereas B and C consist of the potential-like fields.

The actual twist numbers are plotted in the top panel of Figure 5, as a function of the distance, $s$, along the PIL at the nominal height, which is measured from the eastern end of A ($s=0$)  and increases clockwise as denoted in Figure 4. An important limitation in this study is that a twist number as high as 1.75 is hardly found in an FFF model, because of its force-free nature. With this limitation in mind, we note that the most highly twisted field lines ($0.3<T<1.5$) are found in D, which probably belong to the highest possible twist numbers in FFF. The next highest numbers are found in the range of $0.1<T<1$ in A.
We thus regard F2 the most vulnerable to kink instability. It is also noteworthy that the violent eruptions occurred in A and D where $T$ is high.

\begin{figure}[tbh] 
\includegraphics[scale=1.0]{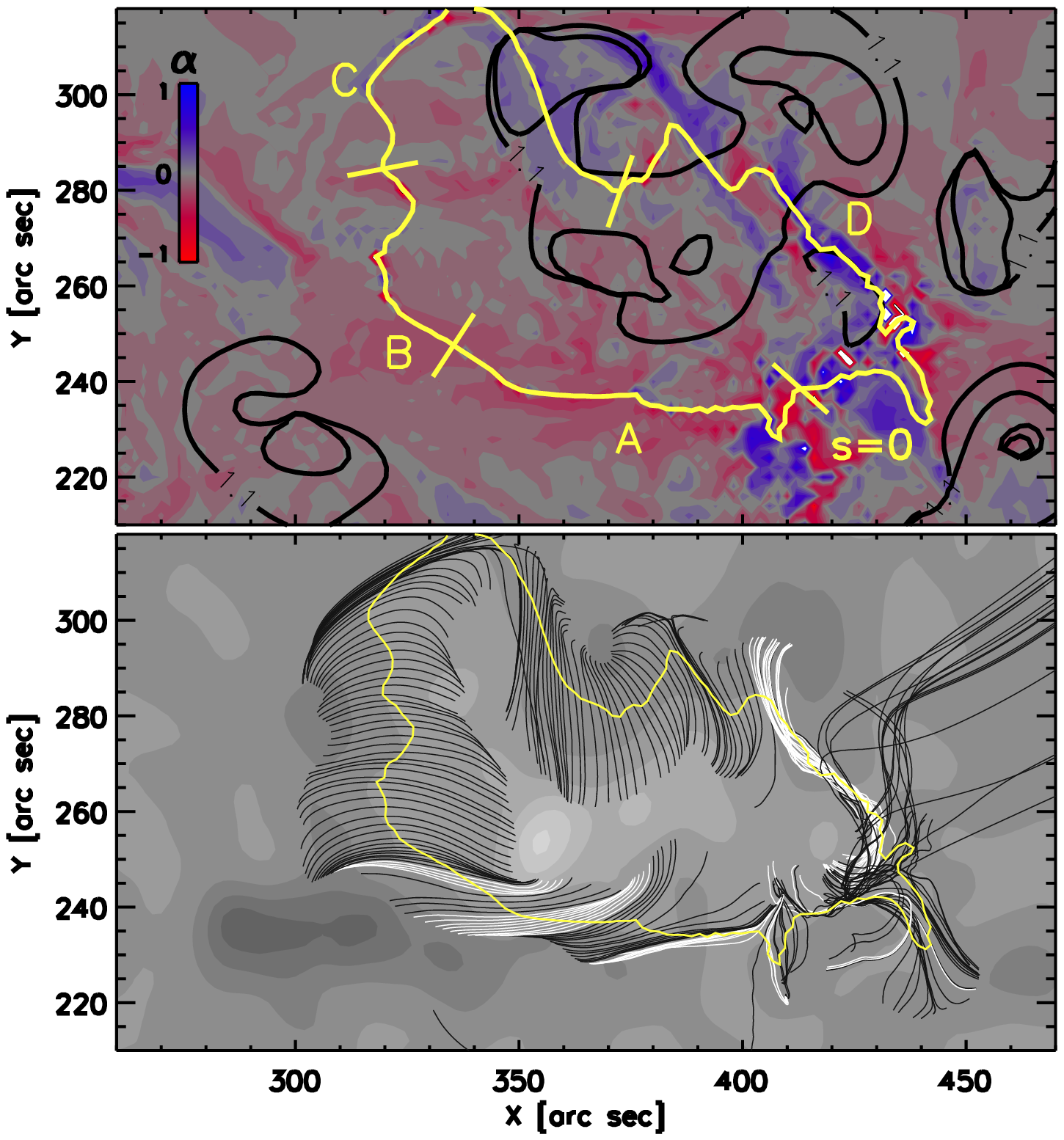}
\caption{Horizontal distributions of the magnetic parameters. {\it Top panel}:  decay index (black contours)  at the height of  $z= 12''$ is shown at two levels, $n=1.1$ and 1.5, on top of the map of $\alpha$ (background image). The yellow contour is the PIL at this height, along which we marked the sections, A--D with a reference point of $s$. {\it Bottom}: field lines passing through the PIL at $z=12''$. Highly twisted field lines with $T\geq 0.5$ are colored white. The background image is the longitudinal magnetic field in the photosphere. }
\end{figure}

\subsection{Torus Instability}

The horizontal distribution of decay index is shown in Figure 4 as black contours at two levels: $n=1.1$ and 1.5. Unlike $\alpha$, the distribution of $n$ shows no particular association with the PIL, along which a filament tends to line up.  The contour of $n=1.5$ appears nowhere above the PIL at the nominal height of the filament, $z=12''$. It is C and D where $n$ reaches the critical value $n_c=1.1$ for torus instability. It is then puzzling how A could host the first eruption and D hosted the last eruption. However, decay index can significantly vary with height, and we need to check its height distribution.

The bottom panel of Figure 5 shows the height distribution of the decay index, $n$, read out along the PIL. The decay index is plotted in the range of $n=-2.1$ (black) to $n=2.1$ (white) in the interval of 0.2. The horizontal lines are the estimated range of the filament height. From  A to the eastern part of C, the decay index shows the usual behavior of monotonic increase with height. In the some part of C and D, however, it shows a reversed variation for a range of height, $50''<z<80''$. Such a variation of $n$  is not uncommon and found to impose no suppression on the process of torus instability.  If $n_c=1.3-1.5$ should apply to this AR, the filament in A should reach $z=23''$ to become unstable to torus instability. On the other hand, the filament in D should reach $z=14''$ only. We can thus expect that F2 should be more vulnerable to torus instability. If even a lower critical decay index, $ n_c=1.1$, is more appropriate to the present magnetic field configuration, the criterion is already satisfied in D, and F2 could have immediately erupted.
{\bf To take into account the uncertainty range of the height estimation, F1 could have already been unstable to torus instability with its location at $z=16''$ (see the upper dashed horizontal line in Figure 5). } In any cases, F2 was more vulnerable to both kink and torus instabilities, and  L1 was more stable than F1 and F2. Nonetheless the eruption occurred in the order of F1, L1, and F2, which is puzzling. We offer a simple explanation for this result {\bf in the next section}.

\begin{figure}[tbh] 
\includegraphics[scale=0.9]{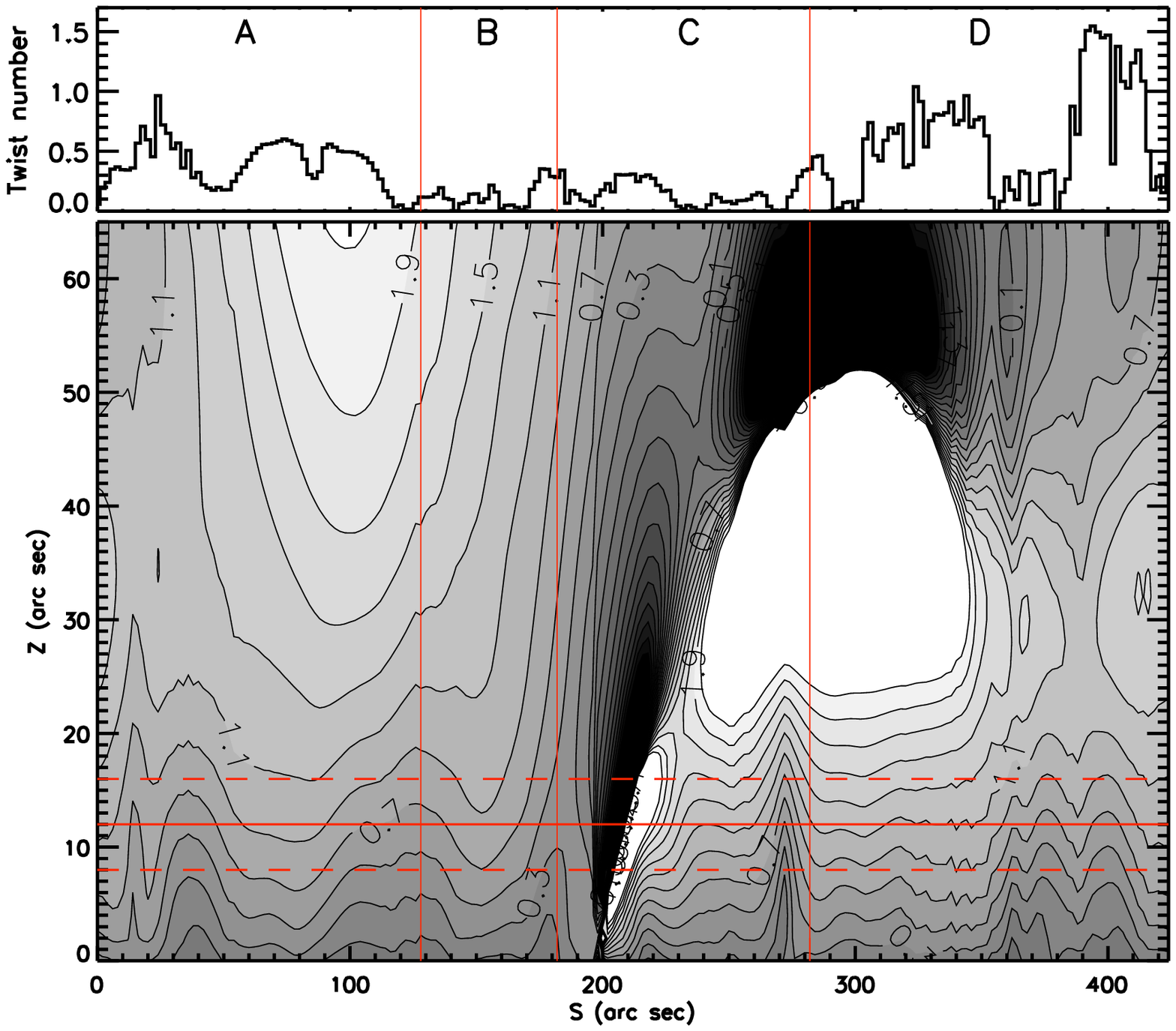}
\caption{Distribution of twist number and decay index along the PIL.  {\it Top}: twist number as a function of the distance, $s$ (yellow contour in Figure 4).
{\it Bottom}: decay index on the $s$-$z$ plane. Only $-2.1\leq n\leq 2.1$ are plotted.  The sections A to D are divided by the vertical red lines. The red solid and dashed horizontal lines are the nominal height of the filaments, $z=12''$ and two additional reference heights, respectively.}
\end{figure}

\section{Discussion and Conclusion}

We have studied the three consecutive eruptions from NOAA AR 11444 using the SDO data, in an attempt to find out what are the main factors determining the likelihood and other behaviors of each eruption. In our results, both the decay index and the twist number show correlations with the eruption speed, in agreement to previous reports by Xu et al. (2012) and Sun et al. (2015). The local-parameter paradigm therefore seems to work to the extent that it can tell how violent an eruption will be, once initiated. However, with the local parameters alone we could not explain which region erupts first, and what determines the time intervals between the eruptions. This leaves open the possibility that some global properties may be responsible for the onset and subsequent evolution of solar eruptions.

Although we did not directly address any global quantities, this study points out a couple of non-local processes associated with the onset of eruptions that have not been considered in the previous studies.
First, {\bf we believe that} the direct cause for the initial eruption was the sporadic heating and plasma motion that led to mass unloading in view of the change of the filament from dark to bright appearance. Since this process was most active in A, F1 could erupt earlier than others. This process occurred over the scale of the AR and the timescale of an hour, which is, therefore, a non-local process. Second, the flare emission expanding toward C evidences that the disturbances created by the eruption in A propagated in the clockwise direction to cause the second eruption in C. Since this AR has a almost round, and closed PIL, the disturbance created by the first eruption could have, in principle, propagated in both directions so that the nearby F2 could erupt earlier than L1.  The unidirectional propagation of the disturbance determined the time sequence of the three successive eruption, which is associated with the overall magnetic structure of the AR, and is not a local property.

An additional question that can be answered under this interpretation is: why there is a latent time in every eruption. It seems that filaments were already near the unstable condition, but did not immediately respond to local perturbations.  For the first eruption, we have witnessed that  none of the tether-cutting like events directly caused the filament to erupt.  The time interval between the tether cuttings and the actual eruption may represent the time taken for the mass unloading, and the latency is due to the gravity in this case. It is yet to be {\bf explored whether the theory of {\bf Myers} et al. (2015) is applicable to the magnetic structure of this active region.} To generalize the implication of the present result, the latency of the eruptions can be understood as the time needed for either the transport of the disturbance created by one eruption site  to the next site, or overcoming the stabilizing force that holds the filament from eruption.

\acknowledgements
We thank anonymous referee for a careful consideration of the manuscript and helpful comments.
This work was supported by the National Research Foundation of Korea (NRF-2012 R1A2A1A 03670387).
J.L. was also supported by the BK21 Plus Program (21A20131111123) funded by the MOE and NRF of Korea,
and especially thanks his hosts at ISEE of Nagoya University for their hospitality.
C.L. is supported by NASA grant NNX13AF76G, NNX14AC12G,  and NNX16AF72G, and
J.J.  by NSF grant AGS-1348513 and NASA grant NNX16AF72G.

\newpage

\end{document}